\documentclass[journal=tosc,draft]{iacrtrans}

\usepackage{tikz}
\usepackage{graphicx}
\usepackage{tikz}
\usepackage{tikz}
\usepackage{multirow}
\usepackage{amsmath}
\usepackage{xcolor}
\usepackage{enumerate}
\usepackage{tabularx}
\newcolumntype{Y}{>{\centering\arraybackslash}X}
\usepackage{booktabs}
\usepackage[titletoc]{appendix}
\usepackage{threeparttable}
\usepackage{algorithm}
\usepackage{algorithmicx}
\usepackage{algpseudocodex}
\usepackage{subfigure}
\usepackage{pdfpages}
\usepackage{color}
\usepackage{colortbl}
\usepackage{float}
\usepackage{nicefrac}
\usepackage{listings}
\usepackage{amssymb}
\usepackage[colorlinks,linkcolor=red]{hyperref}
\newcommand\speck{\textsc{Speck}}

\newcommand\simon{\textsc{Simon}}

\newcommand\etal{\textit{et al.}}

\title{Improving Differential-Neural Distinguisher Model For DES, Chaskey and PRESENT}
\author{Liu Zhang\inst{1}\and
Zilong Wang\inst{1}
}
\authorrunning{Liu Zhang et al.}
\institute{School of Cyber Engineering, Xidian University, Xi'an, China \email{liuzhang@stu.xidian.edu.cn}\\
\email{zlwang@xidian.edu.cn}{}
}

\begin{document}

\maketitle

\keywords{Differential-Neural Distinguisher \and Inception Blocks \and DES \and Chaskey \and PRESENT}

\begin{abstract}
In CRYPTO'19, Gohr proposed a new cryptanalysis strategy using machine learning algorithms.
Combining the differential-neural distinguisher with a differential path and integrating the advanced key recovery procedure, Gohr achieved a 12-round key recovery attack on \textsc{Speck}32/64.
Chen and Yu improved prediction accuracy of differential-neural distinguisher considering derived features from multiple-ciphertext pairs instead of single-ciphertext pairs.
\par
By modifying the kernel size of initial convolutional layer to capture more dimensional information, the prediction accuracy of differential-neural distinguisher can be improved for for three reduced symmetric ciphers.
For DES, we improve the prediction accuracy of (5-6)-round differential-neural distinguisher and train a new 7-round differential-neural distinguisher.
For Chaskey, we improve the prediction accuracy of (3-4)-round differential-neural distinguisher.
For PRESENT, we improve the prediction accuracy of (6-7)-round differential-neural distinguisher.
\footnote{The source codes are available in \url{https://drive.google.com/drive/folders/1i0RciZlGZsEpCyW-wQAy7zzJeOLJNWqL?usp=sharing}.}
\end{abstract}

\section{Introduction}
Classic differential cryptanalysis is a chosen-plaintext attack, which distinguishes ciphertext from random numbers by studying the probability propagation characteristics of specific plaintext differential values in the encryption process, then carries out the key recovery attacks based on differential distinguisher.
The key to classic differential cryptanalysis is to search for a differential distinguisher with a high probability.
Differential-neural cryptanalysis is proposed by Gohr~\cite{crypto/Gohr19} based on classic differential cryptanalysis, using differential-neural distinguisher instead of differential distinguisher.
The differential-neural distinguisher and differential distinguisher have the same role in distinguishing ciphertext from random numbers.
A neural network trains the differential-neural distinguisher as the underlying differential distinguisher, and  Bayesian search is used to speed up key recovery attacks.
If the prediction accuracy of the differential-neural distinguisher is more significantly than 0.5, it is considered an effective distinguisher.
\par
Gohr \cite{crypto/Gohr19} showed that the residual network (ResNet)~\cite{cvpr/HeZRS16} (previously applied in image recognition) could be trained to capture the non-randomness of the distribution of values of output pairs when the input pairs of round-reduced \speck32/64 are of specific difference.
As a result, (5-8)-round (effective) differential-neural distinguishers are trained successfully, and (11-12)-round key recovery attacks for \speck32/64 were achieved by combining 2 rounds of the differential path.
In order to launch more rounds of key recovery attacks, better differential-neural distinguishers were also studied recently.
Chen and Yu~\cite{iacr/ChenY21} proposed multiple-ciphertext pairs instead of single-ciphertext pairs (in Gohr's work) as the input of the neural network. They improved the prediction accuracy of the (5-7)-round differential-neural distinguisher of \speck32/64 to a certain extent.
Bao \etal~\cite{iacr/BaoGLMT21} used Dense Network (DenseNet)~\cite{cvpr/HuangLMW17} and Squeeze-and-Excitation Network (SENet)~\cite{cvpr/HuSS18} with existing deep architectures to train neural network, and  obtained (7-11)-round differential-neural distinguisher and devised a 16-round key recovery attack for \simon32/64.
Zhang et al.~\cite{iacr/zhangWW22} borrowed the idea of the Inception block of GoogLeNet to construct the new neural network architecture. Thus, they trained the differential-neural distinguisher for (5-9)-rounds \speck32/64 and  (7-12)-rounds \simon32/64.
In  EUROCRYPT 2021, Benamira~\cite{eurocrypt/BenamiraGPT21} indicated that Gohr's differential-neural distinguisher builds a good approximation of the differential distribution table of the cipher during the learning phase and learns additional information.
Based on the principle that the purpose of the differential-neural distinguisher is to obtain the difference information in the ciphertext, we have done some tentative work to train a better differential-neural distinguisher on multiple ciphers in this paper. The main improvements for differential-neural distinguisher are listed as follows.
\par
\textbf{Our Contributions.} We modify the network architecture to train differential-neural distinguisher for three reduced symmetric ciphers in this paper.
Compared to Gohr's ~\cite{crypto/Gohr19} and Chen's distinguisher~\cite{iacr/ChenY21}, we improve the prediction accuracy of differential-neural distinguisher for DES, Chaskey, and PRESENT and obtain a more round differential-neural distinguisher for DES under different group size $m$.
\par
The rest of the paper is organized as follows.
Section~\ref{sec:model} introduces the network architecture and train process of our differential-neural distinguisher.
Section~\ref{sec:result} exhibits the prediction accuracy of our differential-neural distinguisher for three reduced symmetric ciphers.
Our work is summarized in Section~\ref{sec:conclusion}.

\section{Our Differential-Neural Distinguisher Model}
\label{sec:model}
Gohr~\cite{crypto/Gohr19} proposed the method of differential-neural cryptanalysis based on classic differential cryptanalysis, where a differential-neural distinguisher is trained using a neural network.
The differential-neural distinguisher is a one-to-many differential path compared to classic differential cryptanalysis.
The input difference of plaintext is identical, but the output difference of ciphertext is different in differential-neural cryptanalysis.
The role of the differential-neural distinguisher is to learn the differential information in the ciphertext.
\subsection{Theoretical Model Function}
The differential-neural distinguisher is a supervised model to distinguish ciphertext and random numbers.
Therefore, it is necessary to construct ciphertext and random numbers artificially, thereby assigning corresponding labels.
Gohr~\cite{crypto/Gohr19} took a single-ciphertext pair as input to the model function, and Chen et al.~\cite{iacr/ChenY21}generalized it, using multiple-ciphertext pairs as input to the model function.
For brevity in the description, two model functions are represented by an expression.
\par
Given $m$ plaintext pairs $\{(P_{i, 0},P_{i, 1})\allowbreak, i \in[0, m-1]\}$ and target cipher, the resulting ciphertext pairs $\{(C_{i, 0},C_{i, 1}),\allowbreak i \in[0, m-1]\}$ is regarded as a sample.
Each sample will be attached with a label $Y$:
\[Y=\left\{\begin{array}{l}
    1, \text { if } P_{j, 0} \oplus P_{j, 1}=\Delta, j \in[0, m-1] \\
    0, \text { if } P_{j, 0} \oplus P_{j, 1} \neq \Delta, j \in[0, m-1]
\end{array}\right.\]
If $Y$ is 1, this sample is sampled from the target distribution and defined as a positive example.
Otherwise, this sample is sampled from a uniform distribution and defined as a negative example.
To guarantee the prediction accuracy of differential-neural distinguisher,
a large number of samples need to be put into neural network training.
If the neural network can obtain a stable prediction accuracy higher than 0.5 on a test set,
it can effectively distinguish ciphertext and random numbers.
The theoretical model function can be described as:
\[\begin{array}{c}
    \operatorname{Pr}(Y=1 \mid X_{0}, \ldots, X_{m-1})=F(f(X_{0}), \ldots, f(X_{m-1}), \varphi(f(X_{0}), \ldots, f(X_{m-1}))) \\
    X_{i}=(C_{i, 0}, C_{i, 1}), i \in[0, m-1] \\
    \operatorname{Pr}(Y=1 \mid X_{0}, \ldots, X_{m-1}) \in[0,1]
\end{array}\]
where $f(X_i)$ represents the basic features of a ciphertext pair $X_i$ and $\varphi(\cdot)$ is the derived features obtained from $f(X_i)$  and $F(\cdot)$ is the new posterior probability estimation function.
In Gohr's model, the value of $m$ is 1~\cite{crypto/Gohr19}. In Chen's model, the value of $m$ is $\{2,4,8,16\}$~\cite{iacr/ChenY21}.
\subsection{Design the Network Architecture}
\begin{figure}[htp]
    \centering
    		\begin{tikzpicture}[help lines/.style={blue!20,very thin} ,scale=0.8, every node/.style={scale=0.8}]

			\filldraw[fill=cyan!20,draw=black,rounded corners] (1,4.5) rectangle (3,5);
			\node at (2,4.75) {Output};
			\filldraw[fill=orange!20,draw=black,rounded corners] (1,5.5) rectangle (3,6);
			\node at (2,5.75) {Module 2};\draw[-latex,thick](2,5.5)--(2,5);
			\fill (2,6.1) circle[radius=1pt];\fill (2,6.25) circle[radius=1pt];\fill (2,6.4) circle[radius=1pt];
			\filldraw[fill=orange!20,draw=black,rounded corners] (1,6.5) rectangle (3,7);
			\node at (2,6.75) {Module 2};

			\filldraw[fill=teal!20,draw=black,rounded corners] (1,7.5) rectangle (3,8);
			\node at (2,7.75) {Module 1};\draw[-latex,thick](2,7.5)--(2,7);
			\filldraw[fill=lightgray!20,draw=black,rounded corners] (1,8.5) rectangle (3,9);
			\node at (2,8.75) {Input};\draw[-latex,thick](2,8.5)--(2,8);

			\draw [draw=pink,thick](0.8,4.35) rectangle(3.2,7.15);
			\draw [draw=pink,thick](0.8,7.35) rectangle(3.2,8.15);
			\node at (0.4,4.75) {$F(\cdot)$};\node at (0.4,7.75) {$f(\cdot)$};

			\filldraw[fill=lightgray!20,draw=black,rounded corners] (3.5,8.5) rectangle (5,9); \node at (4.25,8.75) {Input};\draw[-latex, thick](5.55,8.75)--(5.05,8.75);\node at (6.35,8.75) {$[m,\omega,\frac{2L}{\omega}]$};

			\filldraw[fill=teal!20,draw=black,rounded corners] (3.5,5.75) rectangle (5,6.25); \node at (4.25,6) {Module 1};\draw[-latex, thick](6.45,6)--(5.95,6);

			\filldraw[fill=teal!20,draw=black,rounded corners] (4.25,7.5) rectangle (7.25,8); \node at (5.75,7.75) {$Conv, k_{1}\times k_{1}, N_f$};
			\filldraw[fill=teal!20,draw=black,rounded corners] (7.5,7.5) rectangle (10.5,8); \node at (9,7.75) {$Conv, k_{2}\times k_{2}, N_f$};
			\filldraw[fill=teal!20,draw=black,rounded corners] (10.75,7.5) rectangle (13.75,8); \node at (12.25,7.75) {$Conv, k_{3}\times k_{3}, N_f$};

			\draw[-latex, thick] (5.75,7.5)--(9,6.5);
			\draw[-latex, thick] (9,7.5)--(9,6.5);
			\draw[-latex, thick] (12.5,7.5)--(9,6.5);

			\filldraw[fill=teal!20,draw=black,rounded corners] (7.5,6) rectangle (10.5,6.5); \node at (9,6.25) {$Concatenate$, $3N_f$};
			\filldraw[fill=teal!20,draw=black,rounded corners] (7.5,5.5) rectangle (10.5,6); \node at (9,5.75) {$BN$};
			\filldraw[fill=teal!20,draw=black,rounded corners] (7.5,5) rectangle (10.5,5.5); \node at (9,5.25) {$Relu$};
			\draw[-latex, thick](9,5)--(9,4.5);

			\filldraw[fill=orange!20,draw=black,rounded corners] (0,2) rectangle (2,2.5); \node at (1,2.25) {$k_{s}=k_{s}+2$};

			\filldraw[fill=orange!20,draw=black,rounded corners] (2.25,3.25) rectangle (5.25,3.75); \node at (3.75,3.5) {$Conv,k_s\times k_s, N_f$};
			\filldraw[fill=orange!20,draw=black,rounded corners] (2.25,2.75) rectangle (5.25,3.25); \node at (3.75,3) {$BN$};
			\filldraw[fill=orange!20,draw=black,rounded corners] (2.25,2.25) rectangle (5.25,2.75); \node at (3.75,2.5) {$Relu$};
			\draw[-latex, thick](3.75,4.25)--(3.75,3.75);

			\filldraw[fill=orange!20,draw=black,rounded corners] (2.25,1.75) rectangle (5.25,2.25); \node at (3.75,2) {$Conv, k_s\times k_s, N_f$};
			\filldraw[fill=orange!20,draw=black,rounded corners] (2.25,1.25) rectangle (5.25,1.75); \node at (3.75,1.5) {$BN$};
			\filldraw[fill=orange!20,draw=black,rounded corners] (2.25,0.75) rectangle (5.25,1.25); \node at (3.75,1) {$Relu$};
			\draw[-latex, thick](3.75,0.75)--(3.75,0.35);

			\draw (3.55,0) rectangle (3.95,0.35);
			\node at (3.75,0.15) {$\oplus$};
			\draw[-latex,thick](3.75,0)--(3.75,-0.4);
			\draw[-latex,thick] (3.75,4.1)--(6.5,4.1)--(6.5,0.15)--(3.95,0.15);

			\draw[-latex,thick] (1,2.5)--(1,4.1)--(3.75,4.1);
			\draw[-latex,thick] (3.55,0.15)--(1,0.15)--(1,2);

			\filldraw[fill=orange!20,draw=black,rounded corners] (7.2,2) rectangle (8.7,2.5); \node at (7.95,2.25) {Module 2};
			\draw[-latex, thick](6.6,2.25)--(7.1,2.25);


			\filldraw[fill=cyan!20,draw=black,rounded corners] (8.25,1) rectangle (9.75,1.5); \node at (9,1.25) {Output};
			\draw[-latex, thick](10.35,1.25)--(9.95,1.25);
			\draw[-latex, thick](11.75,2.75)--(11.75,2.25);
			\filldraw[fill=cyan!20,draw=black,rounded corners] (9.75,1.75) rectangle (13.75,2.25); \node at (11.8,1.95) {$GlobalAveragePooling$};
			\filldraw[fill=cyan!20,draw=black,rounded corners] (10.75,1.25) rectangle (12.75,1.75); \node at (11.8,1.5) {$Dropout$};
			\filldraw[fill=cyan!20,draw=black,rounded corners] (10.75,0.75) rectangle (12.75,1.25); \node at (11.8,1) {$Sigmod$};

			\draw[-latex, thick](11.75,0.75)--(11.75,0.25);

		\end{tikzpicture}
    \caption{The network architecture of our differential-neural distinguisher model.}
    \label{fig:distinguisher_model}
\end{figure}
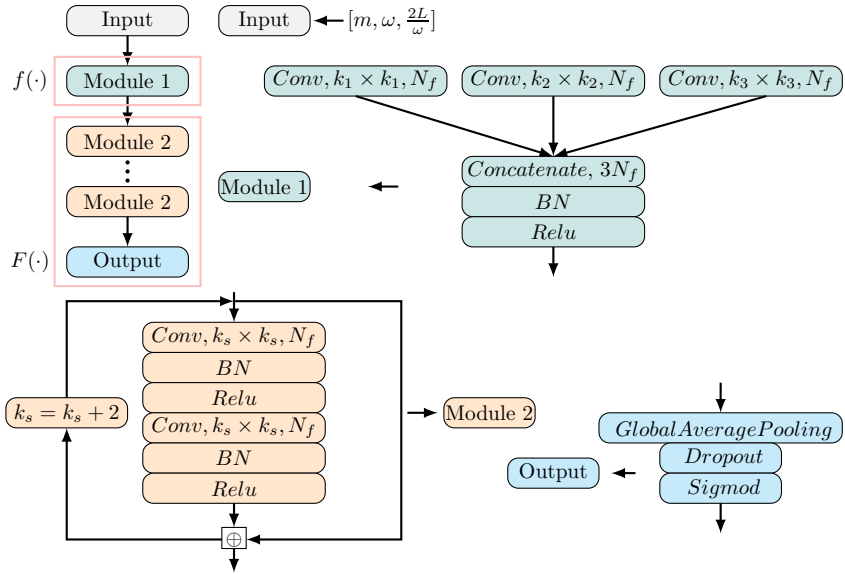

The differential-neural distinguisher is a posterior probability estimation function that evaluates the quality of the distinguisher with prediction accuracy.
Training a differential-neural distinguisher using a neural network is to capture differential information in the ciphertext and unknown information between multiple-ciphertext pairs.
The network architecture of Gohr's~\cite{crypto/Gohr19} and Chen's~\cite{iacr/ChenY21} model mainly includes an initial convolutional layer consisting of width-1 convolutional layers and multiple residual blocks. Zhang et al.~\cite{iacr/zhangWW22} modified the initial convolutional layer using the Inception block instead of the width-1 convolutional layer.
According to theoretical derivation and experiment findings, the following network architecture can ensure the prediction accuracy of the distinguisher to the greatest extent.
The network architecture contains several modules that are described in Figure~\ref{fig:distinguisher_model}.
\par
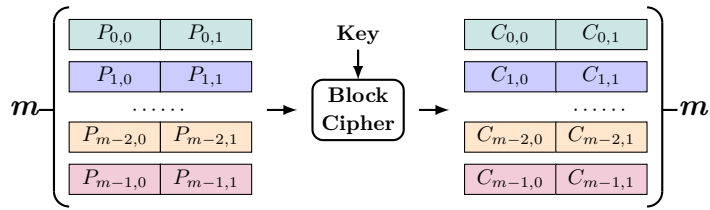
\begin{figure}[htp]
\centering
\begin{tikzpicture}[help lines/.style={blue!50,very thin},scale=0.8, every node/.style={scale=0.8}]

    \filldraw[fill=teal!20,draw=black] (1.5,4.5) rectangle (3,5);
    \node at (2.25,4.75) {$P_{0,0}$};
    \filldraw[fill=teal!20,draw=black] (3,4.5) rectangle (4.5,5);
    \node at (3.75,4.75) {$P_{0,1}$};

    \filldraw[fill=blue!20,draw=black] (1.5,3.8) rectangle (3,4.3);
    \node at (2.25,4.05) {$P_{1,0}$};
    \filldraw[fill=blue!20,draw=black] (3,3.8) rectangle (4.5,4.3);
    \node at (3.75,4.05) {$P_{1,1}$};

    \node at (3,3.5) {$\cdots\cdots$};

    \filldraw[fill=orange!20,draw=black] (1.5,2.8) rectangle (3,3.3);
    \node at (2.25,3.05) {$P_{m-2,0}$};
    \filldraw[fill=orange!20,draw=black] (3,2.8) rectangle (4.5,3.3);
    \node at (3.75,3.05) {$P_{m-2,1}$};

    \filldraw[fill=purple!20,draw=black] (1.5,2.1) rectangle (3,2.6);
    \node at (2.25,2.35) {$P_{m-1,0}$};
    \filldraw[fill=purple!20,draw=black] (3,2.1) rectangle (4.5,2.6);
    \node at (3.75,2.35) {$P_{m-1,1}$};

    \draw [thick](1.5,5.2) arc (90:180:0.25);\draw [thick](1.5,1.9) arc (270:180:0.25);
    \draw [thick] (1.25,4.95) -- (1.25,2.15);
    \draw [thick] (1.25,3.5) -- (1,3.5);
    \node at (0.75,3.525) {$\textit{\textbf{\Large m}}$};

    \draw [-latex, thick] (4.75,3.5) -- (5.25,3.5);


    \draw[rounded corners, thick] (5.5,3) rectangle (7,4);
    \node at (6.25,3.75) {$\textbf{Block}$};
    \node at (6.25,3.25) {$\textbf{Cipher}$};
    \draw [-latex,thick] (6.25,4.5) -- (6.25,4);
    \node at (6.25,4.75) {$\textbf{Key}$};

    \draw [-latex, thick] (7.25,3.5) -- (7.75,3.5);


    \filldraw[fill=teal!20,draw=black] (8,4.5) rectangle (9.5,5);
    \node at (8.75,4.75) {$C_{0,0}$};
    \filldraw[fill=teal!20,draw=black] (9.5,4.5) rectangle (11,5);
    \node at (10.25,4.75) {$C_{0,1}$};

    \filldraw[fill=blue!20,draw=black] (8,3.8) rectangle (9.5,4.3);
    \node at (8.75,4.05) {$C_{1,0}$};
    \filldraw[fill=blue!20,draw=black] (9.5,3.8) rectangle (11,4.3);
    \node at (10.25,4.05) {$C_{1,1}$};

    \node at (10.25,3.5) {$\cdots\cdots$};

    \filldraw[fill=orange!20,draw=black] (8,2.8) rectangle (9.5,3.3);
    \node at (8.75,3.05) {$C_{m-2,0}$};
    \filldraw[fill=orange!20,draw=black] (9.5,2.8) rectangle (11,3.3);
    \node at (10.25,3.05) {$C_{m-2,1}$};

    \filldraw[fill=purple!20,draw=black] (8,2.1) rectangle (9.5,2.6);
    \node at (8.75,2.35) {$C_{m-1,0}$};
    \filldraw[fill=purple!20,draw=black] (9.5,2.1) rectangle (11,2.6);
    \node at (10.25,2.35) {$C_{m-1,1}$};

    \draw [thick](11,5.2) arc (90:0:0.25);\draw [thick](11,1.9) arc (270:360:0.25);
    \draw [thick] (11.25,4.95) -- (11.25,2.15);
    \draw [thick] (11.5,3.5) -- (11.25,3.5);
    \node at (11.75,3.525) {$\textit{\textbf{\Large m}}$};

\end{tikzpicture}
\caption{The arrangement structure of input data}
\label{fig:input}
\end{figure}

\noindent \textbf{Input Module: Data Format.} The neural network receives $m$ ciphertext pairs $\{(C_{i,0},C_{i,1})\ | \ i\in (0,m)\}$ as input data.
We convert a ciphertext pair into a two-dimensional matrix based on the word size of the target cipher.
The input layer of the neural network consisting of multiple-ciphertext pairs is arranged in a $m\times\omega \times \frac{2L}{\omega}$ array, where
$L$ represents the block size of the target cipher, and $\omega$ is the size of a basic unit.
If the target cipher belongs to the Feistel structure, $\omega$ is usually 4.
The generation method and arrangement structure of the input data are shown in Figure~\ref{fig:input}.
\\
\par
\begin{figure}[htp]
\centering
\begin{tikzpicture}[help lines/.style={blue!50,very thin},scale=0.8, every node/.style={scale=0.8}]

    \filldraw[fill=teal!20,draw=black,rounded corners] (4.3,1.5) rectangle (7.7,2); \node at (6,1.75) {$Concatenate,3\times N_f$};
    \filldraw[fill=teal!20,draw=black,rounded corners] (4.3,1) rectangle (7.7,1.5); \node at (6,1.25) {$BN$};
    \filldraw[fill=teal!20,draw=black,rounded corners] (4.3,0.5) rectangle (7.7,1); \node at (6,0.75) {$Relu$};

    \draw[-latex, thick] (2.25,3)--(6,2);
    \draw[-latex, thick] (6,3)--(6,2);
    \draw[-latex, thick] (9.75,3)--(6,2);

    \filldraw[fill=teal!20,draw=black,rounded corners] (0.75,3) rectangle (3.75,3.5);\node at (2.25,3.25) {$Conv,k_{1}\times k_{1},N_f$};
    \filldraw[fill=teal!20,draw=black,rounded corners] (4.5,3) rectangle (7.5,3.5);\node at (6,3.25) {$Conv,k_{2}\times k_{2},N_f$};
    \filldraw[fill=teal!20,draw=black,rounded corners] (8.25,3) rectangle (11.25,3.5);\node at (9.75,3.25) {$Conv,k_{3}\times k_{3},N_f$};

    \draw[-latex, thick] (6,4.5)--(2.25,3.5);
    \draw[-latex, thick] (6,4.5)--(6,3.5);
    \draw[-latex, thick] (6,4.5)--(9.75,3.5);

    \filldraw[fill=teal!20,draw=black,rounded corners] (4.5,4.5) rectangle (7.5,5); \node at (6,4.75) {$\text{Input} \ [m,\omega,N_f]$};

\end{tikzpicture}
\caption{The initial convolution layer}
\label{fig:ini_convo}
\end{figure}
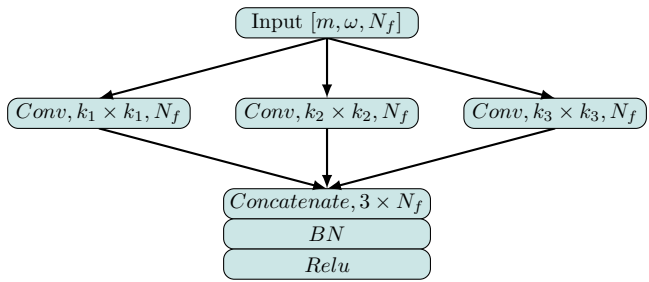

\noindent \textbf{Module 1: Initial Convolution.} After converting the initial ciphertext data to a specific format, the train data enters the initial convolutional layer.
The input layer is connected to the initial convolutional layer, which comprises three convolution layers with $N_f$ channels of different kernel sizes $(k_1,k_2,k_3)$, where ideas come from the Inception block of GoogLeNet.
The three convolution layers are concatenated at the channel dimension.
Batch normalization is applied to the output of concatenate layers.
Finally, rectifier nonlinearity is applied to the output of batch normalization, and the resulting $[m, \omega, 3 \times N_{f}]$ matrix is passed to the Convolutional Blocks layer.
The architecture of the initial convolutional layer can be seen in Figure~\ref{fig:ini_convo}.
\\
\par
\noindent \textbf{Module 2: Convolutional Blocks.} Each convolutional block consists of two convolutional layers of $3\times N_{f}$ filters.
Each block applies first the convolution of kernel size $k_s$, then a batch normalization, and finally a rectifier layer.
At the end of the convolutional block, a skip connection is added to the output of the final rectifier layer of the block to the input of the convolutional block and passes the result to the next block.
After each convolutional block, the kernel size increases by 2.
The amount of convolutional blocks is determined by experiment.
The architecture of convolutional blocks layer can be seen in Figure~\ref{fig:convo_block}.
\begin{figure}[htp]
\centering
\subfigure[The convolutional blocks]{
   \begin{tikzpicture}[help lines/.style={blue!50,very thin},scale=0.8, every node/.style={scale=0.8}]

    \filldraw[fill=orange!20,draw=black,rounded corners] (4.3,3.25) rectangle (7.7,3.75); \node at (6,3.5) {$Conv, k_s\times k_s, 3\times N_f$};
    \filldraw[fill=orange!20,draw=black,rounded corners] (4.3,2.75) rectangle (7.7,3.25); \node at (6,3) {$BN$};
    \filldraw[fill=orange!20,draw=black,rounded corners] (4.3,2.25) rectangle (7.7,2.75); \node at (6,2.5) {$Relu$};
    \draw[-latex, thick](6,2.25)--(6,1.75);
    \draw (5.8,1.35) rectangle (6.2,1.75);
    \node at (6,1.55) {$\oplus$};
    \draw[-latex, thick](6,1.35)--(6,0.85);

    \filldraw[fill=orange!20,draw=black,rounded corners] (4.3,4.75) rectangle (7.7,5.25); \node at (6,4.95) {$Conv, k_s\times k_s, 3\times N_f$};
    \filldraw[fill=orange!20,draw=black,rounded corners] (4.3,4.25) rectangle (7.7,4.75); \node at (6,4.5) {$BN$};
    \filldraw[fill=orange!20,draw=black,rounded corners] (4.3,3.75) rectangle (7.7,4.25); \node at (6,4) {$Relu$};
    \draw[-latex, thick](6,6.25)--(6,5.25);

    \draw[-latex, thick] (6,5.9)--(9,5.9)--(9,1.55)--(6.2,1.55);

    \filldraw[fill=orange!20,draw=black,rounded corners] (2,3.5) rectangle (4,4); \node at (3,3.75) {$k_s=k_s+2$};

    \draw[-latex, thick] (5.8,1.55)--(3,1.55)--(3,3.5);
    \draw[-latex, thick] (3,4)--(3,6)--(6,6);

\end{tikzpicture}
   \label{fig:convo_block}
 } \quad \quad
 \subfigure[The prediction head]{
   \begin{tikzpicture}[help lines/.style={blue!50,very thin},scale=0.8, every node/.style={scale=0.8}]

    \filldraw[fill=cyan!20,draw=black,rounded corners] (1.5,1.5) rectangle (5.5,2); \node at (3.55,1.75) {$GlobalAveragePooling$};
    \filldraw[fill=cyan!20,draw=black,rounded corners] (2.5,1) rectangle (4.5,1.5); \node at (3.55,1.25) {$Dropout$};
    \filldraw[fill=cyan!20,draw=black,rounded corners] (2.5,0.5) rectangle (4.5,1); \node at (3.55,0.75) {$Sigmod$};

    \draw[-latex, thick](3.5,2.5)--(3.5,2);
    \draw[-latex, thick](3.5,0.5)--(3.5,0);

\end{tikzpicture}
   \label{fig:prediction_head}
 }
\end{figure}
\par
\noindent \textbf{Output Module: Prediction Head.} The prediction head consists of a GlobalAveragePooling layer and an output unit using a $Sigmoid$ activation function.
We add a dropout layer (the drop rate is set to 0.8) before the $Sigmoid$ activation function to prevent model overfitting.
The structure of the prediction head is shown in Figure \ref{fig:prediction_head}.
\\
\par
\noindent \textbf{Rationale.}
First, to make it easier for the neural network to capture the differential information of the ciphertext pair, we convert the ciphertext vector into matrices in the input module.
Using multiple-ciphertext pairs with the same distribution as the input of the neural network can significantly reduce the influence of a single misjudgment on the whole result.
Second, the design idea for the initial convolutional layer comes mainly from the Inception block~\cite{cvpr/SzegedyLJSRAEVR15} in GoogLeNet to capture more dimensional information.
In the initial convolutional layer, the size of kernel is $k_1,k_2,k_3$ separately.
In general, we set $k_1$ to 1.
Using the initial width-1 convolutional layer is intended to make learning simple bit-sliced functions easier, such as bitwise addition.
To capture the features of the internal architecture of the cryptographic algorithm, we add the convolution operation of widths $k_2$ and $k_3$ to capture features under different dimensions, such as the circular shift operation, the modular addition operation.
Third, to ensure that the cryptographic algorithm encrypts different rounds without modifying the network architecture, we use a residual network to let the network automatically adjust the model parameters.
In order to capture information in a larger dimension, we modify the size of the residual network convolution kernel in Gohr's model to keep it incrementing 2.
Finally, to prevent the problem of overfitting caused by too few sample sizes, we added a dropout layer.
The essence of using a deep residual network to construct a differential distinguisher is to treat the cipher with a nonlinear round function as a complex function and use multiple residual blocks to fit the function.
\subsection{Model Training Process}
\noindent \textbf{Data Generation.} Training and test data were generated by using the Linux random number generator to obtain uniformly distributed keys $K_{i}$ and plaintext pairs $P_{i}$ with the input difference $\Delta$ as well as a vector of binary-valued real/random labels $Y_{i}$.
During producing training or test data for the target cipher, the plaintext pair $P_{i}$ was then encrypted for $R$ rounds if $Y_{i}=1$, while otherwise, the second plaintext of the pairs was replaced with a freshly generated random plaintext and then encrypted for $R$ rounds.
In this way, training data set $N=10^{7}$ and test data set $M=10^6$ samples were generated for training and testing.
\\
\par
\noindent \textbf{Basic Training Scheme.} We run the training for 20 epochs (denoted by $B_s$) on the dataset of size $10^7$. In order to maximize GPU performance, the batch size (denoted by $E_s$)  processed by the dataset is adjusted according to the parameter $m$.
The last $10^6$ sample was withheld for the test. Optimization was performed against mean square error loss plus a small penalty based on L2 weights regularization parameter $\lambda=10^{-5}$ using the Adam algorithm \cite{corr/KingmaB14}.
A cyclic learning rate schedule was used, setting the learning rate $l_{i}$ for epoch $i$ to $l_{i} = \alpha + \frac{(n-i) \mod (n+1)}{n}.
(\beta-\alpha)$ with $\alpha = 10^{-4},\beta = 2\times 10^{-3}$ and $n=9$.
The networks obtained at the end of each epoch were stored, and the best network by validation loss was evaluated against a test set.
\\
\par
\noindent \textbf{Staged Train Method.} When the number of encryption rounds is large, the basic training scheme described above fails, i.e., the model does not learn to approximate any helpful function.
The staged train method divides the training process of the differential-neural distinguisher into multiple stages.
In~\cite{crypto/Gohr19}, Gohr trained an 8-round distinguisher of \speck32/64 by using the staged train method. For more detailed method details, refer to~\cite{crypto/Gohr19}.

\section{The Experiment Result}
\label{sec:result}
The prediction accuracy is the essential indicator that reflects the performance of the differential-neural distinguisher.
For a cipher reduced to $R$ rounds, a specific plaintext differential is set firstly. Then a training set and test set are randomly generated.
After sufficient training, we will obtain the testing accuracy of the obtained new differential-neural distinguisher.
A key parameter of our differential-neural distinguisher is the number of ciphertext pairs: the group size $m$, which has four options $\{2,4,8,16\}$.
Other parameters related to the training and the network architecture of our differential-neural distinguisher are listed in Table~\ref{tab:paramerter}.
\begin{table}[H]
\centering
\caption{Related parameter for training differential-neural distinguishers}
\label{tab:paramerter}
\begin{tabular}{l@{\hspace{0.5cm}}l@{\hspace{0.5cm}}l@{\hspace{0.5cm}}l}
\toprule
$N_{f}=32$  & $k_{s}=3$ & $B_{s}$ = 1000  & $\lambda = 10^{-5} $      \\
 $l_{i} \in [0.02,0.001] $& $E_{s} = 20$  &  $N = 10^{7} $ &  $M = 10^{6} $ \\
 \bottomrule
\end{tabular}
\end{table}
\par
The baseline distinguisher, abbreviated as $BD$, is reproduced by Chen et al.~\cite{iacr/ChenY21} according to the network architecture of Gohr~\cite{crypto/Gohr19}.
The differential-neural distinguisher of Chen, abbreviated as $MCND$, is trained by using multiple-ciphertext pairs instead of single-ciphertext pairs as the input of the neural network in~\cite{iacr/ChenY21}.
According to the network architecture in Section~\ref{sec:model}, we carried out two sets of experiments.
The $Case_1$ is an experiment using $N=10^7$ samples to train and $M=10^6$ samples to test differential-neural distinguisher.
Also, the $Case_2$ is an experiment using $N=10^7 \times m$ samples to train and $M=10^6 \times m$ samples to test differential-neural distinguisher. Meanwhile, we removed the Dropout layer in the $Case_2$.
According to the structure of the input module in the neural network, the number of multiple-ciphertext pairs in the training set ans test is $\frac{N}{m}$ and $\frac{M}{m}$, respectively.
In $Case_1$, when the value of m is relatively large, the number of samples in the test set will be relatively small, resulting in overfitting, which is why experiment 2 is carried out.
\subsection{Experiments on DES}
\noindent \textbf{Differential-Neural Distinguishers for Reduced DES:} DES~\cite{howard1987data} is a block cipher that is built on a $6\times4$ Sbox. Based on the analysis of DES in~\cite{daglib/0032320}, the plaintext difference adopted in this paper is $\alpha = \text{(0x40080000, 0x04000000)}$ and the baseline distinguishers were built for reduced DES firstly~\cite{crypto/Gohr19}.
Our differential-neural distinguishers are built for DES reduced to 5, 6, and 7 rounds.
The parameter $(k_1,k_2,k_3)$ in the initial convolutional layer are $(1,4,6)$.
The penalty factor is increased to $8\times 10^{-4}$.
Other related parameters are the same as Tabel~\ref{tab:paramerter}.
Corresponding distinguisher prediction accuracy is shown in Table~\ref{tab:des56}.
\begin{table}[htp]
\centering
\caption{Accuracy of differential-neural distinguisher for DES }
\label{tab:des56}
\setlength{\tabcolsep}{0mm}{
\begin{tabularx}{\linewidth}{cYYYYYY}
\toprule
                     & $R$      & $BD$~\cite{crypto/Gohr19}           & $m$=2  & $m$=4  & $m$=8  & $m$=16 \\
\hline
 $MCND$\cite{iacr/ChenY21} & \multirow{3}{*}{5} & \multirow{3}{*}{0.6261}  & 0.7209 & 0.8382 & 0.9318 & 0.9585\\
 $Case_{1}$   &                    & & 0.7206 & 0.8419 & 0.9422 & 0.9831  \\
 $Case_{2}$   &                    & & 0.7224 & 0.8424 & 0.9490 & 0.9939  \\
\hline
 $MCND$\cite{iacr/ChenY21} & \multirow{3}{*}{6} & \multirow{3}{*}{0.5493}  & 0.5653 & 0.5568 & 0.5507 & 0.5532 \\
 $Case_{1}$   &                   &  & $-$    & 0.6135 & 0.6734 & 0.7287 \\
 $Case_{2}$   &                   &  & 0.5728 & 0.6213 & 0.6842 & 0.7603 \\
                   \hline
 $Case_{2}$   & 7                 & $-$ & $-$    & $-$    & 0.5050  & 0.5106 \\
 \bottomrule
\end{tabularx}}
\end{table}
\par
In Table~\ref{tab:des56}, we can see that the distinguisher $BD$ has effectively distinguished ciphertext and random number when the reduced rounds $R = 5$.
Compared to the prediction accuracy of the distinguisher $MCND$~\cite{iacr/ChenY21}, we cannot significantly improve the accuracy of the differential-neural distinguisher where $R=5$.
When $R=6$, the distinguisher $MCND$ cannot significantly improve the prediction accuracy, even if the group size $m$ is constantly increased.
Both in the case of $Case_1$ and $Case_2$, our differential-neural distinguisher significantly increase the prediction accuracy. Meanwhile, we trained successfully the 7-round distinguisher for DES when the group size $m = 8 \text{ and } 16$.
\\
\par
\noindent \textbf{Training 7-round Distinguisher Using the Staged Training Method.} For 7 rounds, the training scheme described above fails, i.e., the model does not learn to approximate any helpful function.
We still succeeded in training a 7-round neural distinguisher of DES by using several stages of pre-training.
First, we use our 6-round distinguisher to recognize 4-round DES with the input difference (0x04000000, 0x40080000) (the most likely difference to appear three rounds after the input difference (0x40080000,0x04000000).
The training was done on $10^7 \times m$ samples for twenty epochs with cyclic learning rates.
Then we trained the distinguisher so obtained to recognize 7-round DES with the input difference $(0x40080000,0x04000000)$ by processing $10^7 \times m$ freshly generated samples for ten epochs with a learning rate of $10^{-4}$.
Finally, the learning rate was dropped to $10^{-5}$ after processing another $10^7 \times m$ fresh samples each.
\subsection{Experiments on Chaskey}
\noindent \textbf{Differential-Neural Distinguishers for Reduced Chaskey:} Based on the best differential path searched in~\cite{sacrypt/MouhaMHWPV14}, baseline distinguishers are built for reduced Chaskey firstly~\cite{crypto/Gohr19}.
Given the plaintext difference $\alpha = (\text{0x8400,0x0400,0,0})$, the baseline distinguisher can distinguish Chaskey up to 4 rounds.
Our differential-neural distinguishers are also built for Chaskey reduced to 3, 4 rounds.
The parameter $(k_1,k_2,k_3)$ in the initial convolutional layer are $(1,5,8)$.
All related parameters are the same with Table~\ref{tab:paramerter}, except for the penalty factor $\lambda$ is increased to $10^{-4}$.
Corresponding distinguisher accuracy is present in Table~\ref{tab:chaskey34}.
\begin{table}[H]
\centering
\caption{Accuracy of differential-neural distinguisher for Chaskey }
\label{tab:chaskey34}
\setlength{\tabcolsep}{0mm}{
\begin{tabularx}{\linewidth}{cYYYYYY}
\toprule
                     & $R$        & $BD$~\cite{crypto/Gohr19}              & $m$=2  & $m$=4  & $m$=8  & $m$=16 \\
\hline
 $MCND$\cite{iacr/ChenY21} & \multirow{3}{*}{3} & \multirow{3}{*}{0.8608}   & 0.8958 & 0.9583 & 0.9887 & 0.9986\\
 $Case_{1}$   &                   &  & 0.9254 & 0.9678 & 0.9892 & 0.9885  \\
 $Case_{2}$   &                  &   & 0.9294 & 0.9820 & 0.9970  & 0.9992  \\
\hline
 $MCND$\cite{iacr/ChenY21} & \multirow{3}{*}{4} & \multirow{3}{*}{0.6161}  & 0.6589 & 0.6981 & 0.7603 & 0.7712 \\
 $Case_{1}$   &                   &  & 0.6755 & 0.7651 & 0.9259 & 0.9403 \\
 $Case_{2}$   &                  &   & 0.6784 & 0.8145 & 0.9309 & 0.9830 \\
 \bottomrule
\end{tabularx}}
\end{table}
\subsection{Experiments on Present}
\noindent \textbf{Differential-Neural Distinguishers for Reduced Present64/80:} Present~\cite{ches/BogdanovKLPPRSV07} is a block cipher that is based on a $4\times 4$ Sbox. Based on the plaintext difference $\alpha = (\text{0,0,0,0x9})$ provide in~\cite{africacrypt/Wang08}, the baseline distinguisher were built for Present64/80 reduced up to 7 rounds~\cite{crypto/Gohr19}.
Our neural distinguishers are also built for Present64/80 reduced to 6, 7 rounds.
The parameter $(k_1,k_2,k_3)$ in the initial convolutional layer are $(1,2,4)$. The concrete parameter of constructing neural distinguisher for Present64/80 are as follows.
Corresponding distinguisher accuracy is present in Table~\ref{tab:present67} .
\begin{table}[H]
\centering
\caption{Accuracy of differential-neural distinguisher for PRESENT }
\label{tab:present67}
\setlength{\tabcolsep}{0mm}{
\begin{tabularx}{\linewidth}{cYYYYYY}
\toprule
                     & $R$        & $BD$~\cite{crypto/Gohr19}             & $m$=2  & $m$=4  & $m$=8  & $m$=16 \\
\hline
 $MCND$\cite{iacr/ChenY21} & \multirow{3}{*}{6} & \multirow{3}{*}{0.6584} & 0.7198 & 0.7953 & 0.8308 & 0.8259\\
 $Case_{1}$   &                   &  & 0.7304 & 0.8116 & 0.8810 &0.9391  \\
 $Case_{2}$   &                  &   & 0.7326 & 0.8204 & 0.9066 & 0.9699 \\
\hline
 $MCND$\cite{iacr/ChenY21} & \multirow{3}{*}{7} & \multirow{3}{*}{0.5486}  & 0.5503 & 0.5853 & 0.5786 & 0.5818 \\
 $Case_{1}$   &                  &   & 0.5717 & 0.6054 & 0.6510 & 0.7070 \\
 $Case_{2}$   &                  &   & 0.5717 & 0.6070 & 0.6559 & 0.7205 \\
 \bottomrule
\end{tabularx}}
\end{table}

\section{Conclusions}
\label{sec:conclusion}
In this article, we modify the network architecture to train differential-neural distinguisher for three reduced symmetric ciphers,
which modify the size of three convolutional kernel in the initial convolutional layer depending on the round function of cipher.
Thus, we improve the prediction accuracy of differential-neural distinguisher and obtained more rounds differential-neural distinguisher for DES, Chaskey, and PRESENT.

\bibliographystyle{alpha}
\bibliography{bib}
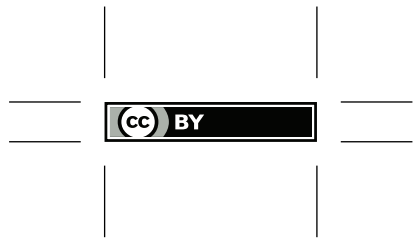
\end{document}